\documentclass[
reprint,
nofootinbib,
superscriptaddress,
amsmath,amssymb,
aps
]{revtex4-2}

\usepackage{graphicx}% Include figure files
\usepackage{dcolumn}% Align table columns on decimal point
\usepackage{bm}% bold math
\usepackage{hyperref}% add hypertext capabilities
\usepackage[version=4]{mhchem}    % For chemical formulas
\usepackage{csquotes}
\hypersetup{colorlinks=true,linkcolor=red,citecolor=blue}
\usepackage{float}
\usepackage{physics}
\usepackage{xspace}
\usepackage[normalem]{ulem}

\usepackage{lipsum}

\usepackage{prettyref}
\newcommand{\pref}[1]{\prettyref{#1}}
\newrefformat{fig}{Fig.~\ref{#1}}
\newrefformat{tab}{Table~\ref{#1}}
\newrefformat{sec}{Sec.~\ref{#1}}
\newrefformat{ssec}{Sec.~\ref{#1}}
\newrefformat{app}{App.~\ref{#1}}
\newrefformat{eqn}{Eq.~(\ref{#1})}

\usepackage{bbding} %checkmarks

\usepackage[dvipsnames,table]{xcolor}

\newcommand{\kcuf}{KCuF$_3$\xspace}
\newcommand{\sfo}{Sr$_2$FeO$_4$\xspace}
\newcommand{\dftu}{DFT+$U$\xspace}
\newcommand{\ulrt}{$U_{\text{LRT}}$\xspace}
\newcommand{\ucrpa}{$U_{\text{cRPA}}$\xspace}

\def\QE{\textsc{Quantum ESPRESSO}\,}

\newcommand{\response}[1]{\textcolor{black}{{#1}}}
\newcommand{\editor}[2]{%
  \expandafter\newcommand\csname #1note\endcsname[1]{%
    \textcolor{#2}{(\textbf{#1:} ##1)}}%
  \expandafter\newcommand\csname #1\endcsname[1]{%
    \textcolor{#2}{##1}}%
  \expandafter\newcommand\csname #1cancel\endcsname[1]{%
    \textcolor{#2}{\sout{##1}}}%
  \expandafter\newcommand\csname #1change\endcsname[2]{%
    \textcolor{#2}{\sout{##1} ##2}}%
  \newenvironment{#1text}{\color{#2}}{\color{black}}
}

\editor{TI}{magenta}

\begin{document}

\title{Bridging constrained random-phase approximation and linear response theory for computing Hubbard parameters}

\author{Alberto Carta} \email{alberto.carta@psi.ch}
\affiliation{Materials Theory, ETH Z\"urich, Wolfgang-Pauli-Strasse 27, 8093 Z\"urich, Switzerland}
\affiliation{PSI Center for Scientific Computing, Theory, and Data, Paul Scherrer Institute, 5232 Villigen PSI, Switzerland}
\author{Iurii Timrov} \email{ iurii.timrov@psi.ch}
\affiliation{PSI Center for Scientific Computing, Theory, and Data, Paul Scherrer Institute, 5232 Villigen PSI, Switzerland}
\author{Sophie Beck}
\affiliation{Center for Computational Quantum Physics, Flatiron Institute, 162 5th Avenue, New York, NY 10010, USA}
\author{Claude Ederer} \email{edererc@ethz.ch}
\affiliation{Materials Theory, ETH Z\"urich, Wolfgang-Pauli-Strasse 27, 8093 Z\"urich, Switzerland}

\date{\today}

\begin{abstract}
The predictive accuracy of popular extensions to density-functional theory (DFT) such as \dftu and DFT plus dynamical mean-field theory (DFT+DMFT) hinges on using realistic values for the screened Coulomb interaction $U$. 
Here, we present a systematic comparison of the two most widely used approaches to compute this parameter, \textit{i.e.} linear response theory (LRT) and the constrained random-phase approximation (cRPA), using a unified framework based on the use of maximally localized Wannier functions.
% We demonstrate good quantitative agreement between LRT and cRPA in cases where the strongly interacting subspace corresponds to an isolated set of bands. Differences can be assigned to neglecting the response of the exchange-correlation potential in cRPA and the presence of additional \response{excitation} channels in LRT.
\response{We show that the $U$ in LRT and cRPA can differ as much as 30\%. We demonstrate that this discrepancy arises from two main differences: neglecting the response of the exchange-correlation potential in cRPA and additional excitation channels in LRT. By taking these differences into account, we can achieve near perfect agreement between the two techniques.}
Moreover, we show that in cases with strong hybridization between interacting and screening subspaces, the application of cRPA becomes ambiguous and can lead to unrealistically small $U$ values, while LRT remains well-behaved.
Our work \response{formally connects} both methods, sheds light on their strengths and limitations, and emphasizes the importance of using a consistent set of Wannier orbitals to ensure transferability of $U$ values between different implementations.
\end{abstract}

% WORDCOUNT
% including abstract we have
% text : 3072 w
% math:  16w * 8 lines = 
% fig1:  300/(0.5*2.3) + 40= 301w # where 2.3 is aspect ratio
% fig2: 300/(0.5/3)+40 = 240w
% total: 3741w including abstract

\maketitle

An accurate description of many quantum materials~\cite{Keimer/Moore:2017} relies on correctly capturing the strong electron-electron interaction among a subset of electrons inhabiting the states around the Fermi level.
The method of choice is to divide the electrons into strongly and weakly interacting subspaces, where the latter can be treated within common approximations to density-functional theory (DFT)~\cite{Hohenberg:1964, Kohn:1965}, while the strongly interacting subspace is augmented with a more accurate treatment of the local electron-electron interaction, resulting in the widely used \dftu~\cite{anisimov_band_1991, Liechtenstein:1995, dudarev_electronenergyloss_1998a, himmetoglu_hubbardcorrected_2013} or DFT plus dynamical mean-field theory (DFT+DMFT) methods~\cite{Georges:1996, kotliar_electronic_2006, held_electronic_2007}.

The accuracy and predictive capabilities of these techniques depend on a realistic estimation of the screened Coulomb interaction within the strongly interacting subspace, which is typically parametrized in terms of an on-site Hubbard $U$. 
% (and potentially an on-site Hund's coupling $J$~\cite{Liechtenstein:1995} and/or an inter-site interaction~\cite{campo_extended_2010}).
Consequently, several first-principles approaches to computing Hubbard parameters have been developed, including constrained DFT~\cite{Dederichs:1984, Mcmahan:1988, Gunnarsson:1989, Hybertsen:1989, Gunnarsson:1990, Pickett:1998, Solovyev:2005, Nakamura:2006, Shishkin:2016}, linear response theory (LRT)~\cite{cococcioni_linear_2005, timrov_hubbard_2018, linscott_role_2018, timrov_selfconsistent_2021, timrov_hp_2022, Macke:2024},  Hartree-Fock-based methods~\cite{Mosey:2007, Mosey:2008, Andriotis:2010, Agapito:2015, TancogneDejean:2020, Lee:2020}, and the constrained random-phase approximation (cRPA)~\cite{Springer:1998, Kotani:2000, aryasetiawan_frequencydependent_2004, aryasetiawan_calculations_2006, Scott2024}.
% \acarta{where to cite the paper requested by the second reviewer?, i have put them here but they are completely irrelevant in my opinion}
%
In particular, LRT and cRPA are widely used in practical applications.
Notably, LRT is used almost exclusively for DFT+$U$ calculations, in particular for high-throughput applications of \dftu~\cite{kirchner-hall_extensive_2021, himmetoglu_hubbardcorrected_2013, capdevila-cortada_performance_2016, moore_highthroughput_2024}, \response{with few exceptions~\cite{Ghosh2023},} while cRPA is utilized more often in the context of DFT+DMFT calculations~\cite{seth_renormalization_2017, souto-casares_oxygen_2021, kazemi-moridani_strontium_2024a}.

% This raises the question to what extent LRT and cRPA can be viewed as equivalent.
\response{Crucially, it has not been established to what extent LRT and cRPA can be viewed as equivalent.}% sophie's formulation
% , \response{especially in light of recent work which has seen LRT $U$ values used in a DFT+DMFT context~\cite{Ghosh2023}}. 
While some attempts were made to compare different  methods for computing Hubbard parameters~\cite{aryasetiawan_calculations_2006, tesch_hubbard_2022, Liu:2023, Yang_2025_arxiv}, there is still lack of understanding and consensus regarding the similarities and differences between them. So far, an explicit quantitative comparison of the calculated $U$ parameters has been hindered by the fact that specific implementations generally use different basis orbitals to represent the strongly interacting subspace, with the $U$ parameter depending very sensitively on the specific choice of these orbitals~\cite{himmetoglu_hubbardcorrected_2013, Carta_et_al:2025, kirchner-hall_extensive_2021, Mahajan:2021}. Many DFT codes implementing the LRT method use internally defined localized orbitals (e.g., orthogonalized atomic orbitals~\cite{Carta_et_al:2025} or partial wave projectors~\cite{Bengone_et_al:2000}),
while cRPA calculations are often based on Wannier functions~\cite{miyake_initio_2009, sakuma_firstprinciples_2013, miyake_forbital_2008, miyake_screened_2008}.

In this letter, we systematically compare $U$ values obtained from LRT and cRPA. We specifically address the need for a consistent definition of the interacting subspace and a consistent choice of \emph{Hubbard projectors}, which represent the corresponding localized basis orbitals. To this end, we extend the LRT calculation of the $U$ parameter using density-functional perturbation theory~\cite{timrov_hubbard_2018, timrov_selfconsistent_2021} within \QE~\cite{giannozzi_quantum_2009a, giannozzi_advanced_2017a, Giannozzi:2020} by interfacing it with \textsc{Wannier90}~\cite{pizzi_wannier90_2020} in order to employ maximally localized Wannier functions (MLWFs)~\cite{Marzari:2012} as Hubbard projectors. 
This enables a direct quantitative comparison between LRT and cRPA, the latter being evaluated using \QE in combination with \textsc{RESPACK}~\cite{nakamura_respack_2021a,kurita_interface_2023}.
\response{We first show mathematically that the definitions of $U$ in the two approaches are not strictly equivalent, but that, for cases where the interacting and screening subspaces correspond to isolated or nearly-isolated sets of bands, one can clearly identify the corresponding differences.
We then demonstrate numerically, that LRT and cRPA can be brought in remarkable quantitative agreement, if these differences are properly accounted for.} 
For cases where the two subspaces hybridize strongly, the application of cRPA is problematic, leading to unrealistically small $U$ values, whereas LRT remains largely unaffected. %\response{And hence, in these cases, the comparison of the two approaches is less straightforward.}

% We begin by briefly summarizing the basic terminology of both approaches.
An essential quantity in both methods is the charge susceptibility, or \emph{response tensor}, which characterizes how a system of interacting electrons reacts to external perturbations. Thereby, one distinguishes between the
\textit{bare} susceptibility, $\chi_0$, which does not consider the change in the electronic interactions caused by the charge rearrangement, and the
\textit{full} susceptibility, $\chi$, which considers both the external perturbation and the effect of the resulting electron redistribution.

\response{Within RPA, the bare susceptibility can be written as a sum over all possible electron-hole excitations in an independent particle basis:
\begin{equation}
\label{eqn:chi-RPA}
\chi_{0}(\omega, \mathbf{r}, \mathbf{r'}) = \sum_{m, n} \frac{f_n - f_m}{\omega + \epsilon_{n} - \epsilon_{m}}
\langle \mathbf{r} | \psi_{m} \rangle \langle \psi_{n} | \mathbf{r} \rangle \langle \mathbf{r'} | \psi_{n} \rangle \langle \psi_{m} | \mathbf{r'} \rangle \,, 
\end{equation}
where $|\psi_n\rangle$, $\epsilon_n$, and $f_n$ are the Kohn-Sham (KS) states, their energies, and occupations, respectively. 
The cRPA formalism~\cite{aryasetiawan_frequencydependent_2004, aryasetiawan_calculations_2006} then emphasizes the division of the electronic Hilbert space into the \textit{interacting subspace}, $\mathcal{D}$, 
% (also called ``target'' subspace or ``correlated'' subspace in the context of DFT+DMFT)
and the \textit{screening subspace}, mirroring the corresponding division within \dftu and DFT+DMFT.~\footnote{In some cases, a different subdivision of the Hilbert space has been used to define the screening subspace within cRPA and to define the interacting subspace orbitals~\cite{miyake_forbital_2008}. 
Here, we always use an identical subdivision to define both screening and interacting subspace.}
% If such a division can be made, one can 
The bare susceptibility is then split into two parts, \mbox{$\chi_0 = \chi_0^\mathcal{D}+\chi_0^\mathcal{R}$}, where $\chi_0^{\mathcal{D}}$ contains only electron-hole excitations within the interacting subspace, and $\chi_0^{\mathcal{R}}$ contains all other transitions.
The partially screened Coulomb interaction acting within the $\mathcal{D}$ subspace is then obtained as
$ W^\mathcal{R}  = \left[ 1  - V \chi_0^\mathcal{R} \right]^{-1} V$, where $V= 1/|\mathbf{r}-\mathbf{r'}|$ is the unscreened Coulomb interaction. 
%
% Alternatively, the Dyson equation can also be formulated in terms of $\chi^{\mathcal{D}}$, the full susceptibility inside the $\mathcal{D}$ subspace, giving $W^\mathcal{R} = (\chi_0^\mathcal{D})^{-1}-(\chi^\mathcal{D})^{-1}$.
For the following it is also useful to define the full susceptibility within the interacting subspace, $\chi^{\mathcal{D}}$, through the Dyson equation $W^\mathcal{R} = (\chi_0^\mathcal{D})^{-1}-(\chi^\mathcal{D})^{-1}$.}

\response{Static interaction parameters are computed as matrix elements of $W^\mathcal{R}(\omega=0)$ with the Hubbard projector functions $\phi^I_i(\mathbf{r})$, {\it i.e.}, the specific basis orbitals used to represent $\mathcal{D}$, corresponding to orbital $i$ on site $I$~\cite{aryasetiawan_frequencydependent_2004}:}
\begin{align}
&(W^\mathcal{R})_{i j k l}^{I J K L}=\int\int \phi_{i}^{I}({\bf{r}})^{*}\phi_{j}^{J}({\bf{r}})\nonumber \\ 
&\times W^\mathcal{R}(\omega=0, \mathbf{r},\mathbf{r'})
\phi_{k}^{K}({\bf{r}}^{\prime})^{*}\phi_{l}^{L}({\bf{r}}^{\prime})\ d{\bf{r}}\ d{\bf{r}}^{\prime} .
    \label{eqn:crpa_static_elements}
\end{align}
%
% For practical calculations, one usually defines the Hubbard $U$ and the Hund's exchange parameter $J$ by suitably averaging over matrix elements belonging to a single site.
\response{Within the Slater parametrization of the local interaction, which is often employed in 
%DFT+$U$ 
calculations using the full $d$ shell~\cite{Liechtenstein:1995}, $U$ is defined as average over both inter- and intra-orbital matrix elements (see, \textit{e.g.}, Ref.~\cite{merkel_calculation_2024a}): 
\begin{align}
    U_{\text{cRPA}}^I  &= \frac{1}{(N^I)^2} \sum_{ij} (W^\mathcal{R})^{IIII}_{iijj} 
    \label{eqn:definition_U_cRPA} \ , %\quad\quad % \\ 
    % U^I_{\text{cRPA}}-J^I_{\text{cRPA}}  &= \frac{1}{M(M-1)} \sum_{i j} \left[ (W^\mathcal{R})^{IIII}_{iijj} -(W^\mathcal{R})^{IIII}_{ijji} \right] .
    % \label{eqn:definition_U_minus_J_cRPA}
\end{align}
where $N^I$ is the number of interacting orbitals on site $I$.}
% ~\footnote{\response{Note that DFT+DMFT calculations sometimes employ the so-called Kanamori parametrization, where $U$ is defined as average of the interaction tensor over only the intra-orbital matrix elements, i.e., as $\tfrac{1}{M} \sum_i (W^\mathcal{R})^{IIII}_{iiii}$, which is generally larger than $U^I_\text{cRPA}$ defined in \pref{eqn:definition_U_cRPA}.}}
% by about $\tfrac{8}{7}J^I_\text{cRPA}$~\cite{kazemi-moridani_strontium_2024}. 
%
% \response{We also remark} that LRT is typically discussed in the context of a further simplified form of the interaction tensor introduced by Dudarev \textit{et al.}~\cite{dudarev_electronenergyloss_1998a}, 
% with a single effective Hubbard parameter which can be identified as $U_{\text{eff}} = U - J$, with $U$ and $J$ corresponding to the Slater parametrization.
% Accordingly, in this work, we will compare $U_{\text{LRT}}$ with both $U_{\text{cRPA}}$ as well as with $U_{\text{cRPA}} - J_{\text{cRPA}}$.

\response{On the other hand, in LRT, the relevant susceptibilities are obtained by directly calculating the response of the KS system to small localized perturbations, with $\tilde{\chi}^{IJ} = \partial n^I/\partial \lambda^J$ defined in terms of the change in occupation of the Hubbard projector orbitals on site $I$ due to a local potential shift on site $J$, $\lambda^J$, acting only within the interacting subspace.
The Hubbard parameter for site $I$ is then defined as the corresponding diagonal element of the difference of the inverse susceptibility matrices~\cite{himmetoglu_hubbardcorrected_2013, cococcioni_linear_2005}:  
\begin{equation}
    U^I_{\text{LRT}}  =(\tilde{\chi}_0^{-1} - \tilde{\chi}^{-1})^{II} \ ,
    \label{eqn:definition_U_linear_response}
\end{equation}
where $\tilde{\chi}_0$ is calculated without considering the change in the KS potential resulting from the charge rearrangement. Note that $\tilde{\chi}_0$ and $\tilde{\chi}$ describe the 
%``coarse grained'' 
density-density response only on a per site basis, without any orbital dependence, i.e., they describe a purely \emph{monopolar} response.
However, as shown in Ref.~\cite{himmetoglu_hubbardcorrected_2013}, these ``coarse-grained'' susceptibilities can be related to a more general, site- and orbital-dependent, 4-index response function.}
\response{In Sec.~I.F of the supplemental information (SI)~\cite{SI} we show that for cases where the interacting subspace $\mathcal{D}$ is formed from an isolated set of bands, the cRPA susceptibilities $\chi_0^\mathcal{D}$ and $\chi^\mathcal{D}$ can be identified with this general 4-index response function if one neglects the exchange-correlation (xc) contribution to the susceptibility (the latter is included in LRT but typically neglected in cRPA).}
% % by summing over all orbital channels on the corresponding sites, i.e, 
\response{The coarse grained response can then be written in terms of the orbital averages of the bi-diagonal components, $\tilde{\chi}_0^{IJ} = \sum_{ij}(\chi_0^\mathcal{D})_{iijj}^{IIJJ}$ and $\tilde{\chi}^{IJ} =\sum_{ij} (\chi^\mathcal{D})_{iijj}^{IIJJ}$~\cite{SI}.}
\begin{figure*}
   \centering
   \includegraphics[width=\textwidth]{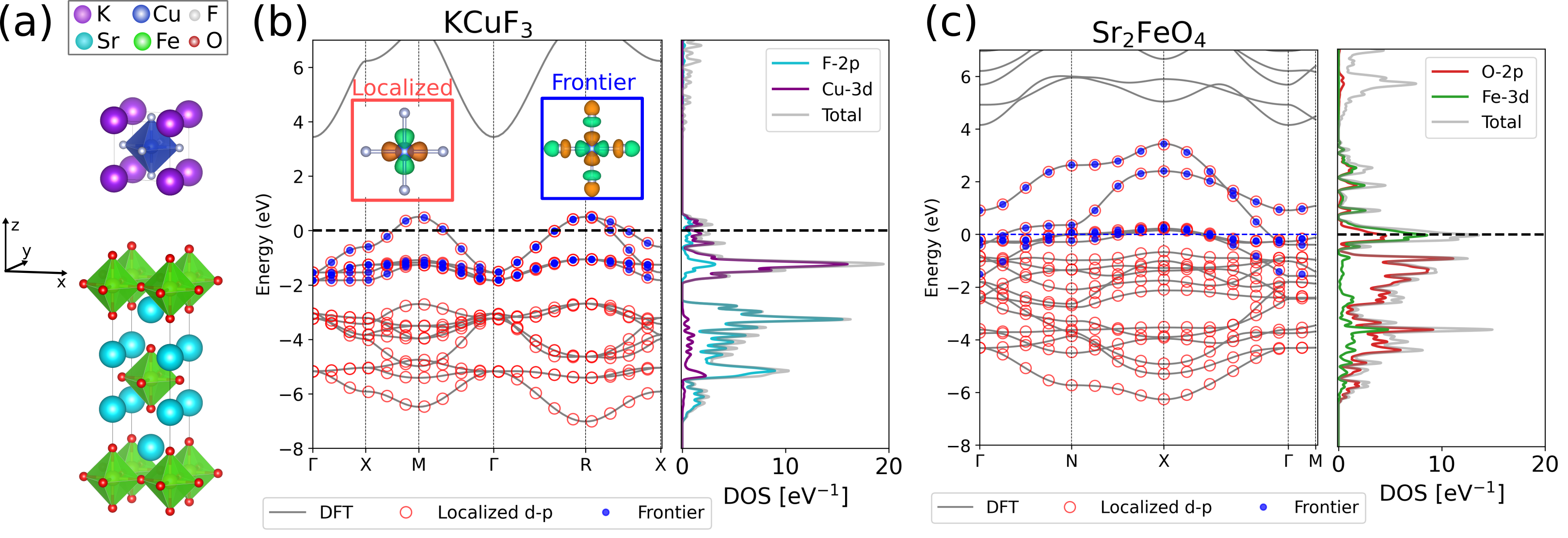}
   \caption{(a) Crystal structures of cubic \kcuf  and tetragonal \sfo.
   (b) and (c) Band structure and projected densities of states (DOS) of (b) \kcuf and (c) \sfo. Zero energy corresponds to the Fermi level. Open and closed circles indicate the bands recalculated from the localized $d$-$p$ and frontier orbital basis sets, respectively. The red and blue insets in (b) show isosurface plots of the Cu $d_{x^2-y^2}$ MLWF in the localized $d$-$p$ and frontier orbital basis sets, respectively. 
   \label{fig:materials_bands}
   }
\end{figure*}
% \response{
% % \sdb{In Sec.~I.F of the supplemental information (SI)~\cite{SI} we show that for cases where the interacting subspace $\mathcal{D}$ is formed from an isolated set of bands, $\chi_0$ and $\chi$...}
% % Putting all together, we prove mathematically in Sec.~I.F of the supplemental information (SI)~\cite{SI}, for cases where the interacting subspace $\mathcal{D}$ is formed from an isolated set of bands, $\chi_0$ and $\chi$ can be identified with the cRPA susceptibilities $\chi_0^\mathcal{D}$ and $\chi^\mathcal{D}$ if one neglects the exchange-correlation (xc) contribution to the full susceptibility. }
% % the site-resolved coarse grained response can be related to the 4-index response within $\mathcal{D}$ by summing over all the orbital responses on the site
% % $\tilde{\chi}_0^{IJ} = \sum_{ij}(\chi^\mathcal{D}_0)_{iijj}^{IIJJ}$ and $\tilde{\chi}^{IJ} =\sum_{ij} (\chi^\mathcal{D})_{iijj}^{IIJJ}$  (see SI~\cite{SI} for more details).}
% %
% %
\response{$U_{\text{LRT}}$ can then be related to the cRPA interaction tensor, \pref{eqn:crpa_static_elements}, through:
\begin{multline}
\label{eqn:main_text_expliciting_U_tilde_2}
U_{\text{LRT}}^{I} = \sum^{ZT}_{zt}\sum^{OPQS}_{opqs}(\tilde{\chi}_0^{-1})^{IZ}  \\ \times
\left[({\chi^\mathcal{D}_0})_{zzpo}^{ZZPO} (W^\mathcal{R})_{opqs}^{OPQS} ({\chi^\mathcal{D}})_{sqtt}^{SQTT} \right](\tilde{\chi}^{-1})^{TI} \ ,
% \TI{= (\tilde{\chi}_0^{-1})^{II} - (\tilde{\chi}^{-1})^{II}}
\end{multline}
% Here, $\chi_0$ and $\chi$ refer to the generalized susceptibilities, expressed in the orbital basis corresponding to the interacting subspace, while $\tilde{\chi}_0$ and $\tilde{\chi}$ refer to the coarse-grained LRT susceptibility matrices, with $\tilde{\chi}^{IR}=\sum_{ir}\chi^{IIRR}_{iirr}$~\cite{himmetoglu_hubbardcorrected_2013}.
%
where all indices correspond to $\mathcal{D}$.}

% %
\response{If we separate the cRPA susceptibilities into their locally ``isotropic'' component (i.e., the purely monopolar part that gives rise to $\tilde{\chi}_0$ and $\tilde{\chi}$) and a locally ``non-isotropic'' part,
% ($\Delta\chi^\mathcal{D}_0$,$\Delta\chi^\mathcal{D}$),
see Eq.~(42) in the SI~\cite{SI}, the isotropic contributions cancel with the corresponding coarse-grained susceptibilities in \pref{eqn:main_text_expliciting_U_tilde_2}, resulting in the orbitally-averaged $U_\mathrm{cRPA}^{I}$ from \pref{eqn:definition_U_cRPA} plus all remaining terms containing non-isotropic parts. 
Thus, we can write \mbox{$U_\text{LRT}^{I} = U_\text{cRPA}^{I} + \Delta U^{I}$},
% The orbitally averaged $U_\mathrm{cRPA}^{I}$ arises from the cancellation of the isotropic components of the susceptibility, which therefore do not participate in the screening.
where $\Delta U^{I}$ contains all non-isotropic contributions. These additional terms correspond to excitations completely inside the interacting subspace, and are thus excluded from the screening within cRPA. They represent processes that alter the ``shape'' of the charge density on an atomic site (orbital reorganizations, bond fluctuations) and also include local exchange terms containing matrix elements of the form $(W^\mathcal{R})^{OOOO}_{oppo}$.}

\response{Thus, in the case of well separated interacting and screening subspaces, we can identify two differences in the definition of the $U$ parameter between  LRT and cRPA: (1) the xc contribution to the response is typically neglected in cRPA but included in LRT, and (2) the coarse-grained nature of LRT introduces additional terms $\Delta U^I$ related to non-isotropic excitations within $\mathcal{D}$.}

We \response{now investigate how these differences manifest numerically within realistic calculations, using two different materials as benchmarks: \kcuf and \sfo. Sec.~V of the SI~\cite{SI} contains results for two additional materials: NiO and CrO$_2$.}

\kcuf has been studied using both DFT+$U$ and DFT+DMFT~\cite{Liechtenstein:1995, Pavarini/Koch/Lichtenstein:2008}, due to its Jahn-Teller-distorted structure and associated orbital order. For the purpose of this work, we consider \kcuf in the high symmetry cubic perovskite structure [shown in \pref{fig:materials_bands}(a)], both for ease of computation and since it exhibits a prototypical band structure of a transition metal (TM) perovskite, with a clear separation between bands with dominant Cu $d$ and F $p$ character [see \pref{fig:materials_bands}(b)]. 
\sfo crystallizes in a layered perovskite structure [shown in \pref{fig:materials_bands}(a)]~\cite{dann_structure_1993, dann_synthesis_1991}, and is of interest due to its structural and electronic similarity to the unconventional superconductor Sr$_2$RuO$_4$~\cite{kazemi-moridani_strontium_2024a}. The corresponding KS bands [see \pref{fig:materials_bands}(c)] exhibit a small overlap between bands with dominant Fe $d$ and dominant O $p$ character. As reported recently, the computed value of $U_\text{cRPA}$ for the Fe $d$ states is too small to reproduce the experimentally observed insulating state of \sfo within DFT+DMFT~\cite{kazemi-moridani_strontium_2024a}.  

\begin{figure*}
   \centering
   \includegraphics[width=0.9\textwidth]{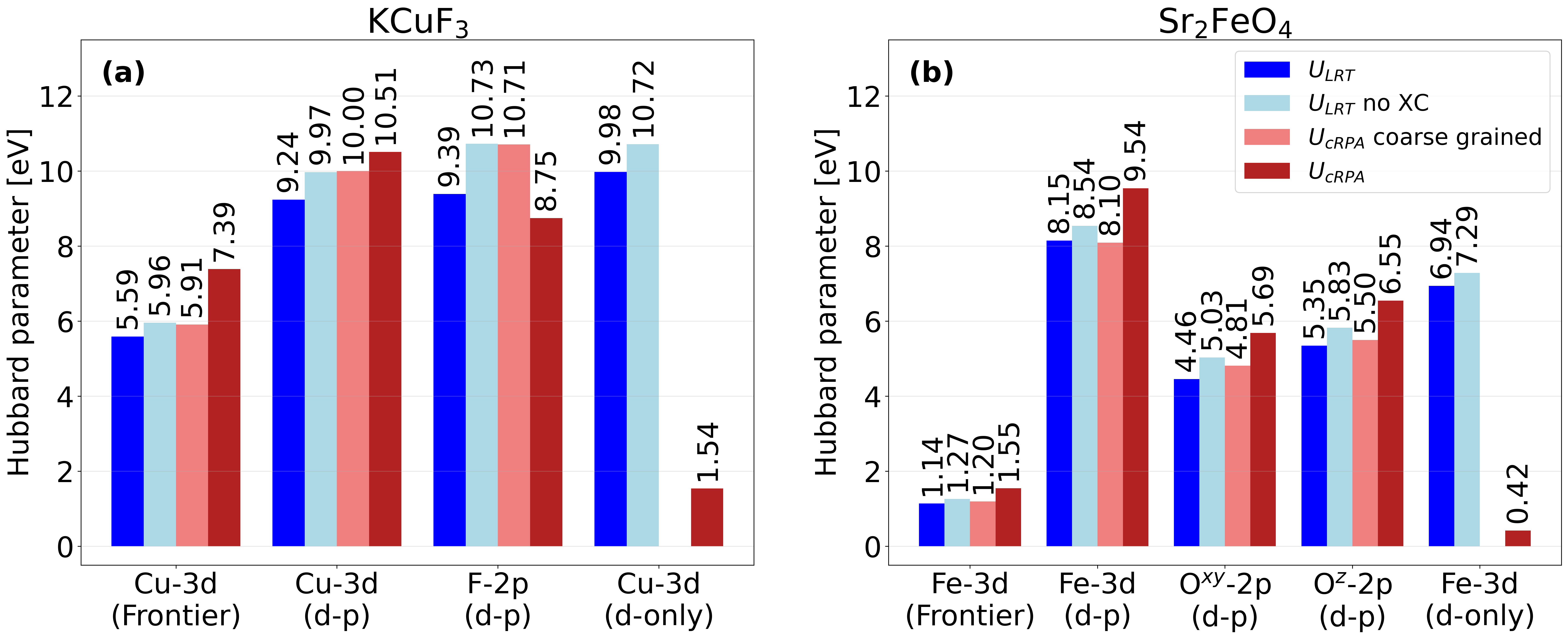}
   \caption{Interaction parameters obtained from LRT and cRPA for (a) \kcuf and (b) \sfo using different choices for the Hubbard projectors \response{ (frontier, $d$-$p$, and $d$-only). Dark blue and dark red bars correspond to interaction parameters computed from a standard LRT and cRPA calculation respectively. The light blue bar corresponds to a modified LRT calculation, i.e. by removing the contribution of the xc kernel, while the light red bar corresponds to the modified cRPA calculation, i.e. obtained by using the coarse graining.}
   For \sfo, O$^{xy}$ refers to the oxygen within the Fe planes, while O$^{z}$ refers to the apical oxygen. \response{The coarse-grained cRPA results for the $d$-only model are not physically meaningful and are thus not included here.}
     \label{fig:summary_plot}
   }
\end{figure*}

For both materials, we compare three different choices of the interacting subspace. In each case we represent the corresponding basis orbitals as MLWFs constructed from the KS band stuctures shown in \pref{fig:materials_bands}, obtained for the nonmagnetic, {\it i.e.}, spin-degenerate, state. 
In the first case, we consider all bands with dominant TM $d$ and ligand $p$ character [see the projected densities of states in \pref{fig:materials_bands}(b) and (c)], which are completely separated from other bands at lower and higher energies for both materials, and construct a complete MLWF basis for these bands consisting of five TM centered orbitals and three ligand centered orbitals per F or O within the unit cell. The resulting MLWFs closely resemble atomic $d$ and $p$ orbitals, repectively [see inset in \pref{fig:materials_bands}(b) for the example of the Cu $d_{x^2-y^2}$-like orbital in \kcuf]. We call this the \emph{localized $d$-$p$} basis.
In the second case, we consider the same set of bands but construct only five MLWFs centered at the TM sites.
These MLWFs again closely resemble atomic $d$ orbitals but are slightly more localized compared to the $d$-$p$ basis. We call this second case {\it localized $d$-only}. 
Finally, in the third case, we restrict the energy window for the calculation of the MLWFs to include only the ``frontier'' bands with dominant TM $d$ character immediately around the Fermi level. 
This results in five more delocalized $d$-like MLWFs, centered at the TM sites but exhibiting $p$-like tails at the surrounding ligand sites, which reflects the hybridized character of the corresponding bands [see the blue inset of \pref{fig:materials_bands}(b) for the Cu $d_{x^2-y^2}$-like orbital in \kcuf]. We call this the \emph{frontier orbital} basis. Such orbitals are often used in DFT+DMFT calculations.

In \pref{fig:summary_plot}, we summarize the calculated values for various definitions of $U$ for both \kcuf(a) and \sfo(b) and the three different choices for the interacting subspace.
We first focus on the localized $d$-$p$ and the frontier orbital basis, which completely span the KS bands within the respective energy windows [see \pref{fig:materials_bands} (b) and (c)].
\response{The interaction parameters obtained from LRT (shown in dark blue) are of comparable magnitude to those obtained from cRPA (in dark red), with \ucrpa being larger than \ulrt by around 10-30\%. The only exception are the F $p$ states in \kcuf in the $d$-$p$ basis, for which \ulrt is about 7\% larger.} This last result can be explained by the known tendency of LRT to overestimate $U$ in cases with essentially completely full (or completely empty) shells, where the applicability of the linear response approach becomes problematic~\cite{tesch_hubbard_2022, yu_communication_2014, himmetoglu_hubbardcorrected_2013}. Indeed, the occupation of the F $p$ states in \kcuf is nearly \response{1.0} electrons per orbital
in our $d$-$p$ basis, while in \sfo, the stronger hybridization with the TM $d$ states reduces the average occupation of the O $p$ states to $\sim$ 0.9, such that the applicability of LRT is not problematic.
% Best agreement is typically observed between $U_{\text{LRT}}$ and $U_{\text{cRPA}}-J_{\text{cRPA}}$, which supports the notion that $U_\text{LRT}$ should be viewed as effective $U-J$.
%
% The agreement is particularly good for the localized $d$-$p$ basis, for which $U_{\text{LRT}}$ and $U_{\text{cRPA}}-J_{\text{cRPA}}$ differ by less than $3\%$, except for the F site in \kcuf. For the frontier case the difference is somewhat larger, around $12\%$.

\response{To relate these quantitative differences  between $U_\text{LRT}$ and $U_\text{cRPA}$ to our mathematical considerations, we first analyze the effect of removing the response of the xc potential from the LRT calculation (light blue bars in \pref{fig:summary_plot}), as explained in Sec.~II in the SI~\cite{SI}. This generally increases $U_{\text{LRT}}$ by $5$-$10\%$, thus bringing it closer to the cRPA value. To analyze the remaining difference to $U_\text{cRPA}$, we then perform a coarse-graining of the cRPA results by computing $\tilde{\chi}_0$ and $\tilde{\chi}$ from the 4-index tensors $\chi^\mathcal{D}_0$ and $\chi^\mathcal{D}$ (see Sec.~I in the SI~\cite{SI}), and extract a corresponding $U$ via \pref{eqn:definition_U_linear_response}. The resulting value (light red bars in \pref{fig:summary_plot}) is mathematically the same quantity as \ulrt with the xc response neglected. Indeed, the corresponding values show very good quantitative agreement. They are nearly identical for \kcuf, while for \sfo they exhibit residual discrepancies of about $5\%$.
These small numerical differences can be assigned to the finite convergence threshold used in the calculations and the fact that RESPACK computes, in addition to all density-density terms, $(W^\mathcal{R})^{IIJJ}_{iijj}$ across all cells, only the exchange integrals, $(W^\mathcal{R})^{IJJI}_{ijji}$, in the first unit cell. Other elements are thus missing from our coarse-graining procedure.
% \sdb{does it not have to do with the fact it's not a truly isolated set of bands?} \acarta{}
% since we do not enforce any symmetry between Wannier functions)
% and/or \acarta{the neglecting of terms} when computing the cRPA interaction tensor (e.g., RESPACK computes all density-density terms $(W^\mathcal{R})^{IIJJ}_{iijj}$, but, except for the exchange integrals $(W^\mathcal{R})^{IJJI}_{ijji}$ in the first unit cell, it neglects all other $W^\mathcal{R}$ matrix elements). 
Overall, the remarkable quantitative agreement between modified LRT and cRPA for both the frontier and $d$-$p$ bases, despite using completely different implementations, strongly supports our analytical result.}

% The overall good quantitative agreement between LRT and cRPA for the frontier and $d$-$p$ subspaces indicate that, at least for the materials considered here, these differences are small.
% , in particular if the local exchange terms in \pref{eqn:main_text_expliciting_U_tilde_2} are taken into account in an effective way by considering $U_\text{cRPA}-J_\text{cRPA}$. 
%
% However, we also note that test calculations where we removed the exchange-correlation part of the response within LRT led to an increase of $U_\text{LRT}$ by about 5-10\%, depending on the specific case. This shows that both this effect as well as the additional terms in \pref{eqn:main_text_expliciting_U_tilde_2} can contribute notably to the difference between $U_\text{LRT}$ and $U_\text{cRPA}$.

We now turn to the localized $d$-only case (last set of bars in both panels of \pref{fig:summary_plot}), for which \response{the interacting and screening subspaces are strongly hybridized.}
Several methods have been suggested to obtain $\chi_0^\mathcal{R}$ in such cases~\cite{miyake_initio_2009, sasioglu_effective_2011, Kaltak:2015, Kaltak_spectral_2025}. Here, we use the \textit{weighted method} introduced in Ref.~\cite{sasioglu_effective_2011}.
%
% In this case LRT and cRPA  lead to drastically different results.
While \ucrpa is very small in this case, \ulrt is much larger and comparable to the values obtained for the localized $d$-$p$ basis.
% The very small values for $U_\text{cRPA}$
% % and $U_\text{cRPA}-J_\text{cRPA}$
% obtained for the $d$-only case are related to the strong hybridization between the interacting and screening subspaces in the underlying bands.
% %
% % , which assigns weights to each particle-hole transition in \pref{eqn:chi-RPA} according to the product of the weights of the $\mathcal{D}$ subspace in the corresponding Bloch functions. 
% %
\response{The very small $U_\text{cRPA}$ for the $d$-only case is due to the strong screening  from transitions with very low energy, related to bands with mixed character that are close to or even crossing the Fermi level.}
%
% this weighted method can \response{still include}  transitions with very low energy which contribute strongly to the screening and thus lead to a very small $U_\text{cRPA}$.% and a resulting overscreening.}
%
% This behavior is closely related to the general tendency of cRPA to overscreen~\cite{vanloon_random_2021, honerkamp_limitations_2018, werner_dynamical_2016, shinaoka_accuracy_2015}, which is strongest if the interacting and screening subspaces are close in energy, such that corrections beyond RPA become important~\cite{vanloon_random_2021}. 
%
%
The $d$-only case thus represents a notoriously problematic case, where the general applicability of the cRPA approach becomes questionable, andthe results can depend sensitively on the specific way the two subspaces are separated.~\footnote{We note that 
even though the ``projector method''~\cite{Kaltak:2015} (not implemented in RESPACK) typically results in larger interaction parameters compared to the weighted method (see, e.g., \cite{merkel_calculation_2024a, Reddy/Kaltak/Kim:2025}), it nevertheless suffers from the same problem~\cite{merkel_calculation_2024a}.}
On the other hand, the $d$-only case is quite relevant for practical applications, since in DFT+$U$ and DFT+DMFT calculations for TM compounds, one typically applies the $U$ correction only on the TM sites, while the ligand states are considered as weakly interacting.

Within LRT, apart from the fact that the $d$-only MLWFs are slightly more localized, the main difference to the $d$-$p$ case is the absence of TM--ligand off-diagonal elements in the susceptibility matrices $\tilde{\chi}^{IJ}$ and $\tilde{\chi}_0^{IJ}$, which affect $U^I_\text{LRT}$ due to the inversion in \pref{eqn:definition_U_linear_response}.
These off-diagonal elements are larger in \sfo than in \kcuf, due to the stronger $d$-$p$ hybridization in the oxide compared to the more ionic fluoride,  which results in a larger difference of $U_{\text{LRT}}$ between the two localized basis sets for \sfo.

We note that, since the generalized susceptibility $(\chi_0)_{ijkl}^{IJKL}$ is equivalent to the bare cRPA susceptibility~\cite{SI} (for clearly separable $\mathcal{D}$), the \response{well known overscreening effects identified for cRPA when interacting and screening subspaces are close in energy or overlap}~\cite{vanloon_random_2021, honerkamp_limitations_2018a, werner_dynamical_2016, shinaoka_accuracy_2015, Chang2024}, can be expected to also affect $U_\text{LRT}$ in such cases. \response{As noted recently, such overscreening can also be due to an incorrect energy separation between the frontier and ligand states in the underlying band structure ~\cite{Carta/Panda/Ederer:2025}.}
% even if the practical problem of how to separate two strongly hybridized subspaces does not appear explicitly. 
This could explain why in the \sfo frontier basis, for which $\mathcal{D}$ overlaps slightly with the lower-lying screening bands, \ulrt and \ucrpa  are too low to reproduce the experimentally observed insulating state~\cite{kazemi-moridani_strontium_2024a}. 

In summary, we have incorporated the use of Wannier functions as Hubbard projectors in LRT calculation of the Hubbard $U$, enabling a direct quantitative comparison with corresponding cRPA calculations. 
\response{Our analysis shows that, in cases where the interacting subspace corresponds to an isolated set of bands, both approaches can be understood within the same framework.  Differences in the corresponding $U$ values are related to neglecting the xc contribution to the response within cRPA and the coarse-graining to purely monopolar response functions within LRT, which effectively incorporates additional excitation channels within the interacting subspace.
Our numerical results show that, if these differences are appropriately taken into account, excellent quantitative agreement can be achieved, thus solving the long standing question about the equivalence of the two methods.}
% in cases where the interacting subspace corresponds to an isolated set of bands, \response{both LRT and cRPA are based on similar approximations and show good quantitative agreement.}  
% % We further showed, that the quantitative agreement can be understood also on a formal level, since both methods are based on similar approximations.
% Differences \response{between standard definitions of \ulrt and \ucrpa} arise from neglecting the change of the exchange-correlation potential in the cRPA response and the fact that LRT takes into account additional screening effects related to local exchange as well as inter-site and inter-orbital contributions within the interacting subspace.
For cases where the interacting and screening subspaces do not correspond to a separated set of bands, we find that cRPA produces unphysically small interaction values, questioning its applicability in such cases.

% While $U_\text{LRT}$ appears unaffected by this problem, the fact that LRT does not systematically exclude screening within the interacting subspace could lead to a ``double-counting'' in the subsequent treatment of interaction effects within the interacting subspace.
% \acarta{Instead of the paragraph above should we have a word about U-J and refer to the SI?} \CE{Not sure. Now we never mention $U-J$ anywhere in the text, so it would be weird to suddenly start discussing it in the summary. Unless perhaps we can present it as some sort of "outlook".}
%
\response{In practice, given the close relation between the two methods, we believe that choosing one over the other depends mostly on what kind of calculation (DFT+$U$, DFT+DMFT, etc.) is of interest and whether an explicit treatment of the effects coming from the xc contribution or the exclusion of the multipolar screening channels is considered to be important.}
% \acarta{A sentence recovering the fact that we have solved a long standing problem}
% \acarta{I think there is value in a statement like: when this is not important either method is fine}
% \sdb{I think the question that needs to be addressed here (taken from the cover letter) is how "our results, [which] resolve a fundamental methodological division". The previous sentence reads like in the end we're saying that the two communities can carry on using their two distinct methods - we've checked and the math works out. While this may be the result, one needs to phrase this carefully to not undermine your work}

Finally, we point out that using well-defined Wannier projectors not only allows for a systematic comparison between LRT and cRPA (and potentially other methods to calculate $U$), but also offers greater transferability across different implementations. This can help to remedy the large spread in reported $U$ values used for the same materials but obtained with different DFT codes and potentially different subspace definitions.

We are thankful to Matteo Cococcioni and Kazuma Nakamura for valuable discussions. 
This research was supported by ETH Z\"urich. The Flatiron Institute is a division of the Simons Foundation. I.T. acknowledges partial support by the NCCR MARVEL, a National Centre of Competence in Research, funded by the Swiss National Science Foundation (SNSF) Grant No.~205602, and SNSF Grants No.~200021-227641 and No.~200021-236507. Calculations were performed on the \enquote{Euler} cluster of ETH Z\"urich.

\bibliography{alberto_zotero, other_references}% Produces the bibliography via BibTeX.
\end{document}